# Optically Pumped Magnetometry in Arbitrarily Oriented Magnetic Fields

S. J. Ingleby, I. C. Chalmers, C. O'Dwyer, P. F. Griffin, A. S. Arnold, and E. Riis
Department of Physics, SUPA, Strathclyde University, 107 Rottenrow East, Glasgow, UK
Email: stuart.ingleby@strath.ac.uk

*Abstract*—Optically pumped atomic magnetometers (OPMs) offer highly sensitive magnetic measurements using compact hardware, offering new possibilities for practical precision sensors. Double-resonance OPM operation is well suited to unshielded magnetometry, due to high sensor dynamic range. However, sensor response is highly anisotropic with variation in the orientation of the magnetic field. We present data quantifying these effects and discuss implications for the design of practical sensors.

## I. Introduction

Alkali vapour cell magnetometry utilises optical pumping and probing to achieve precise magnetic field measurements through detection of magnetisation evolution in a polarised atomic sample. The use of coherent light sources has facilitated improvement in the sensitivity of laboratory devices to aT level [1], competitive with superconducting quantum interference device (SQUID) magnetometers [2]. Unlike SQUIDs, optically pumped magnetometers do not require cryogenic temperatures, making the technique well suited for the design of compact sensors [3].

The highest OPM sensitivities have been achieved using atomic samples operating in the spin exchange relaxation free (SERF) regime [4], in which atomic decoherence is suppressed in a highly polarised, high pressure atomic vapour. This technique allows miniaturised sensors to be realised using micro-fabricated atomic vapour cells, and significant advances have been made in the development of this technology [5-6]. However, SERF-based magnetometers only enjoy narrow magnetic resonances in low (< 10 nT) magnetic fields, and are rarely operated outside high-permeability magnetic shielding [7]. Unshielded sensor operation is desirable in a wide range of practical magnetometer applications, from surveying [8] to geophysics [9], involving measurements of magnetic fields in the geophysical range (~50 μT). Double-resonance magnetometry, in which the response of an optically-pumped atomic sample to an oscillating perturbation is measured, allows sensors to be designed with high dynamic range, well-suited to geophysical measurements [10]. We discuss the practical application of a double-resonance technique to unshielded magnetometry, with measurements of sensitivity and signal response in a laboratory test system.

This work is funded by the UK Quantum Technology Hub in Sensing and Metrology, EPSRC (EP/M013294/1).

## II. Double-resonance Magnetometry

The Larmor frequency $\omega_L$ of atomic spin precession in a polarised sample is proportional to the magnetic field magnitude, and can be measured by detecting a resonant response to an oscillating perturbation $\omega_{RF} \sim \omega_L$. The sample may be perturbed by modulation of the pump light amplitude [11], frequency [12] or polarisation [13], or by a small applied field $B_{RF}$ [14]. On resonance $\omega_L = \omega_{RF}$, a large modulation response is observed in the absorption or polarisation of light transmitted through the sample. The magnetic resonance linewidth is dependent on the polarisation and coherence time of the sample atomic spins, which are optically pumped using coherent light resonant with an atomic hyperfine transition. The simultaneous exploitation of both optical resonance (light resonant with the hyperfine transition, for pumping and probing the sample) and magnetic resonance (RF modulation resonant with Zeeman level splitting) is the basis of the double-resonance magnetometry technique.

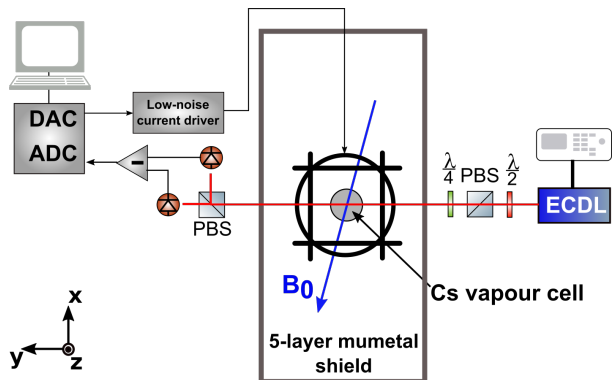

Figure 1: Schematic showing experimental system, including extended-cavity diode laser (ECDL), polarisation optics, digital-to-analogue (DAC) and analogue-to-digital (ADC) converters.

The laboratory test system used for shielded sensor characterisation is shown in Figure 1. The $^{133}$Cs atomic vapour sample is contained within a low-pressure, antirelaxation-coated glass cell [15], and is optically pumped and probed by monochromatic light resonant with the $6^2S_{1/2}$ (F = 4) to $6^2P_{1/2}$ (F = 3) transition. A small (< 5 nT) modulation field $B_{RF}$ is applied on the z-axis using a Helmholtz coil pair. By monitoring the polarisation modulation of light transmitted through the sample, the resonant atomic spin response may be detected. Demodulation of this signal with reference to the applied field modulation yields a Lorentzian resonance line shape centred on the Larmor frequency with linewidth $\Gamma$.

This system configuration has been chosen in order to develop techniques for compact devices. A single monochromatic pump/probe laser beam is used in order to minimise optical hardware requirements – for a compact sensor the ECDL shown in the test system may be replaced by a vertical cavity surface emitting laser (VCSEL). The atomic sample may also be contained within a microfabricated thermal vapour cell [5-6]. Figure 2 shows the configuration of a portable sensor implementing these features. A miniaturised polarimeter system, incorporating polarisation optics, photodiodes and differential preamplifier is used. The shielding and three-axis compensation coils are used in the test system to characterise sensor response in precisely controlled static field magnitudes and orientations [16].

By contrast with SERF magnetometry, use of a double-resonance technique does impose an additional overhead in signal processing and demodulation. A practical system operating in the geophysical range requires a low complexity signal processing chain operating at a few MHz.

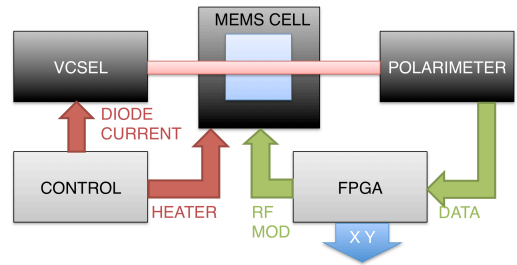

Figure 2: Schematic showing configuration of portable double-resonance magnetic sensor. Millimeter-scale MEMS-fabricated alkali vapour cells may be used, alongside minature VCSEL lasers and compact firmware-based modulation/demodulation (FPGA).

Such a system is well within the capabilities of modern FPGAs/DSPs and can be implemented using inexpensive, compact hardware. The test system performs modulation generation, signal demodulation and line shape analysis in software.

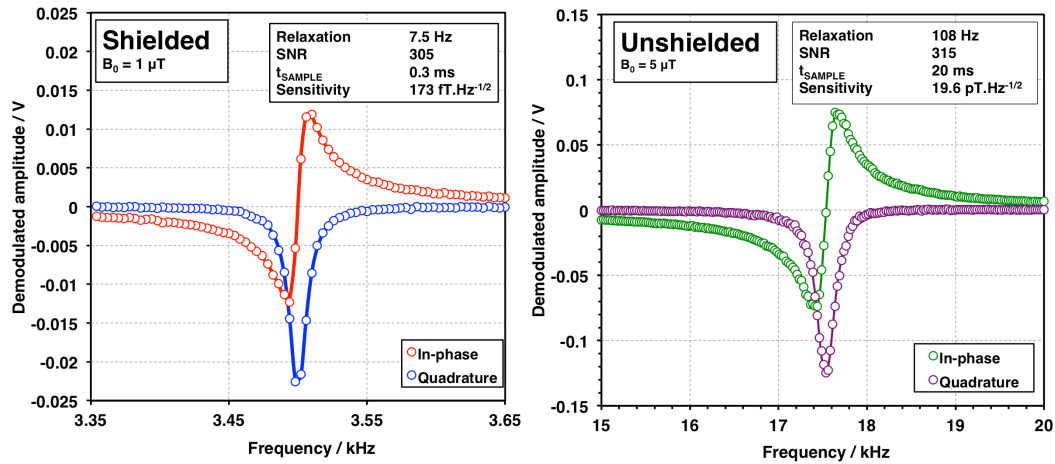

Figure 3: Magnetic resonance data recorded in both shielded and unshielded field measurements. The linewidth of the magnetic resonance is assumed to be equal to the atomic spin relaxation rate, and the signal-to-noise ratio (SNR) estimated by residuals to the fitted resonance curve. The presence of AC line noise limits the unshielded sampling rate to 50 Hz.

Figure 3 shows magnetic resonances measured using both shielded and unshielded test systems. A conservative estimate is shown for magnetometric sensitivity, based on the approximation that sensor resolution is limited by an intrinsic noise source of flat intensity over the sensor bandwidth.

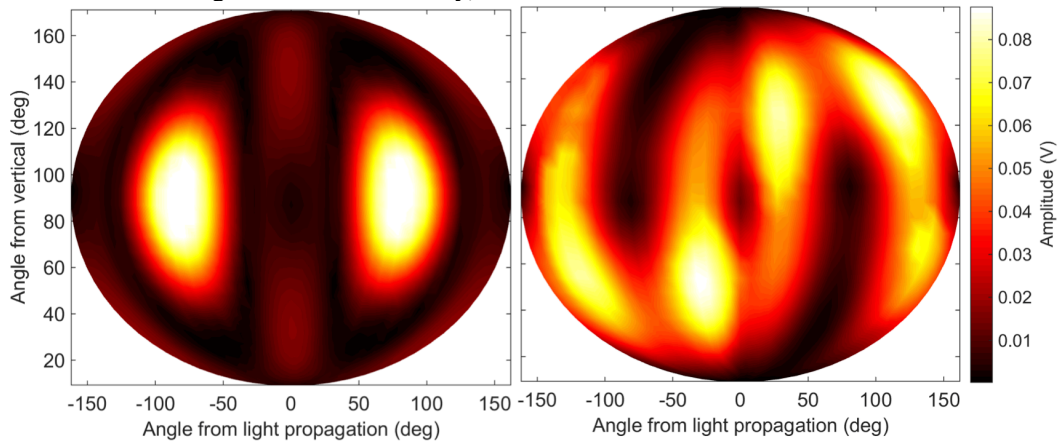

Figure 4: Measured distribution of on-resonance signal amplitude with varying static field orientation over full 4π solid angle. Left: linearly polarised pump-probe light; right: elliptically polarised pump-probe light.

## III. SENSOR RESPONSE ANISOTROPY

A double-resonance magnetometer used for unshielded measurements may be placed in a static field of arbitrary or unknown orientation. Because the evolution of induced orientation and alignment in the atomic polarisation is geometry-dependent, the amplitude, phase and linewidth of the resonance signal response are highly dependent on the static field orientation relative to the pump light propagation and polarisation [17-18]. The precise field control of the test system allows these effects to be measured and characterised. Figure 4 shows the distribution of signal response amplitude under varying static field orientation and pump light polarisation. Figure 5 shows the distribution of magnetic resonance linewidth under varying static field orientation.

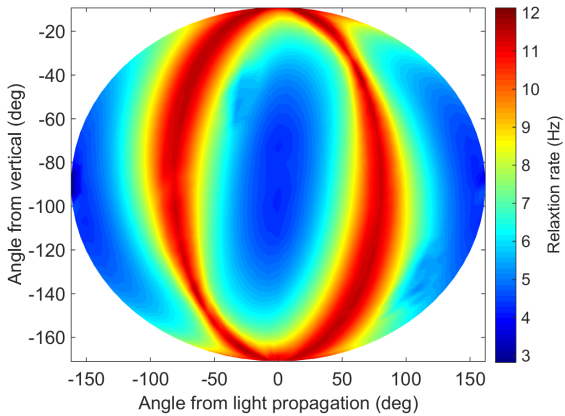

**Figure 5: Magnetic resonance linewidth distribution measured over full 4π angular distribution of static field orientation.**

A full treatment of anisotropic static field effects, including a theoretical model and further experimental data is given in [19].

## CONCLUSIONS

Double-resonance magnetometry offers a technique for development of compact, precise sensors, with sensitivity and dynamic range suitable for enhancing a wide range of unshielded magnetometry applications. The sensor configuration described uses optical hardware economically, relying on digital signal processing to achieve high sensitivity and bandwidth. This approach is well suited to scalable, portable devices. The test system described has been used for precise measurement of the strong dependence of signal amplitude upon light polarisation and static field orientation. Using this information, optimised sensor configurations may be designed for specific measurement applications.

## ACKNOWLEDGEMENTS

The authors would like to thank Prof. Antoine Weis and Dr. Victor Lebedev of Fribourg University for supplying the Cs vapour cells used in this work.